\documentclass[prl,twocolumn,showpacs,groupedaddress]{revtex4}

\usepackage{graphicx}

\begin{document}

\bibliographystyle{apsrev}

\def\be{\begin{equation}} 
\def\ee{\end{equation}} 
\def\bee{\begin{eqnarray}} 
\def\eee{\end{eqnarray}} 
\def\sech{\mbox{sech}}
\def\e{{\rm e}} 
\def\d{{\rm d}} 
\def\kb{k_{\rm B}}
\def\tw{t_{\rm w}}
\def\ts{t_{\rm s}}
\def\Tc{T_{\rm c}}
\def\gs{\gamma_{\rm s}}
\def\tm{tunneling model }
\def\TM{tunneling model }
\def\tilde{\widetilde}
\def\Deltac{\Delta_{0\rm c}}
\def\Deltamin{\Delta_{0\rm min}}
\def\Emin{E_{\rm min}}
\def\tauc{\tau_{\rm c}}
\def\tauac{\tau_{\rm ac}}
\def\tauw{\tau_{\rm w}}
\def\taumin{\tau_{\rm min}}
\def\taumax{\tau_{\rm max}}
\def\de{\delta\varepsilon / \varepsilon}
\def\pF{{\bf pF}}
\def\pFAC{{\bf pF}_{\rm ac}}
\def\ie{i$.$e$.$ }
\def\Ie{I$.$e$.$ }
\def\eg{e$.$g$.$ }
\def\Eg{E$.$g$.$ }
\def\fig{Fig$.$~\ref}
\def\Fig{Fig$.$~\ref}
\def\figs{Figs$.$~\ref}
\def\Figs{Figs$.$~\ref}
\def\sec{section~\ref}
\def\Sec{Section~\ref}
\def\eq#1{(\ref{#1})}
\title{Field--induced structural aging in glasses at ultra low temperatures}

\author{S. Ludwig and D. D. Osheroff}
\affiliation{Department of Physics, Stanford University, Stanford, California
  94305-4060, USA}

\date{\today}

\begin{abstract}

In non--equilibrium experiments on the glasses Mylar and BK7, we measured the
excess dielectric response after the temporary application of a strong electric
bias field at mK--temperatures. A model recently developed
describes the observed long time decays qualitatively for Mylar~\cite{Lud03b},
but fails for BK7. In contrast, our results on both samples can be described by
including an additional mechanism to the mentioned model with temperature
independent decay times of the excess dielectric response. 
As the origin of this novel process beyond the "tunneling model" we suggest bias
field induced structural rearrangements of "tunneling states" that decay by
quantum mechanical tunneling.
\end{abstract}


\maketitle

Most low temperature properties of glasses are well described by the
phenomenological "tunneling model"~\cite{Zel71,And72,Phi72,Phi87}.
It starts from the assumption that the potential minima of groups of atoms are
not "well defined", but can be described as particles moving in "double well"
potentials. At low enough temperatures, only the ground states of the two
wells are occupied, and the "tunneling states" (TSs) move by quantum mechanical
tunneling between them. The energy splitting of these two levels is given by
$E=\sqrt{\Delta_0^2+\Delta^2}$, where the asymmetry, $\Delta$, is the energy
difference between the two local minima. The tunneling splitting can be
estimated using the WKB--approximation as
$\Delta_0\simeq\hbar\Omega\exp(-\lambda)$, with the attempt frequency,
$\Omega$. The tunnel parameter $\lambda=d\sqrt{2mV}/2\hbar$ contains the
effective mass, $m$, the potential barrier height, $V$,
and the distance between the two wells, $d$. Recent experiments revealed
deviations from the tunneling model and show that
interactions between TSs gain importance with decreasing
temperature~\cite{Ens02,OshDip,Lud03b}. Our new results will be discussed
within a picture that goes beyond "double well" potentials.

Long time non--equilibrium dynamics of TSs in the polyester glass Mylar
after the temporary application of a large electric bias field
were previously presented~\cite{Lud03b}.
These data were analyzed using the "dipole gap" theory~\cite{Bur95JLTP}
by extending its predictions to the case of a temporary applied bias field.
A central assumption of the "dipole gap" theory is that mutually strongly
coupled TSs do not contribute to the dielectric response in thermal equilibrium.
However, out of equilibrium these TSs
contribute to the dielectric constant as if isolated~\cite{Bur95JLTP}. 
The findings in~\cite{Lud03b} were explained in terms of strongly coupled
pairs of TSs that are driven out of equilibrium during a field sweep, not only
by relaxation, but also by non--adiabatic driving of
one member of two strongly coupled TS--pair by the applied field. 
Our new results indicate the existence of
an additional process leading to non--equilibrium pairs of TSs,
while the sum of all processes remains constant.
This novel process yields an excess dielectric response with a
decay time independent of the bias field history and temperature.
For the polyester glass Mylar, the new process becomes significant only at a
sufficiently high electric bias field.
In our experiments, we measured the dielectric constant of a $15~\mu$m
thick Mylar film and a $300~\mu$m thick BK7 wafer, both contained in a
$^3$He immersion cell. All measurements were performed in the linear
response region and at the frequency $f=10$~kHz.

\begin{figure}[th]
\includegraphics[width=0.95\linewidth]{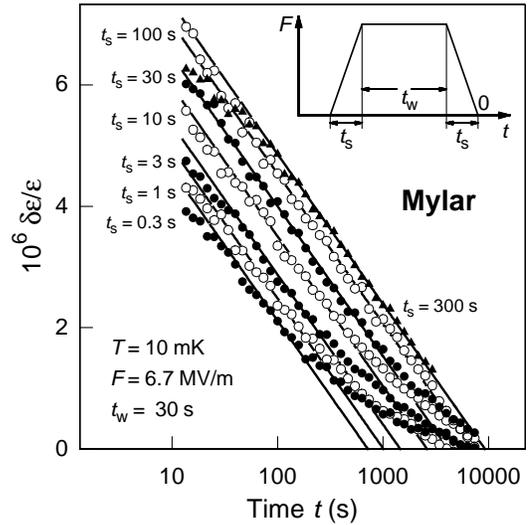}
\vskip -8mm 
\caption{Dielectric response of Mylar to a temporary applied bias field with
maximal value $F$ for a waiting time $\tw=30$~s and different sweep times
at a temperature $T=10$~mK. Solid lines are model curves explained in the
text.  The schematic illustrates the bias field sweep and the definition of
$t=0$.}
\label{fig1}
\end{figure}

\begin{figure}[th]
\includegraphics[width=0.95\linewidth]{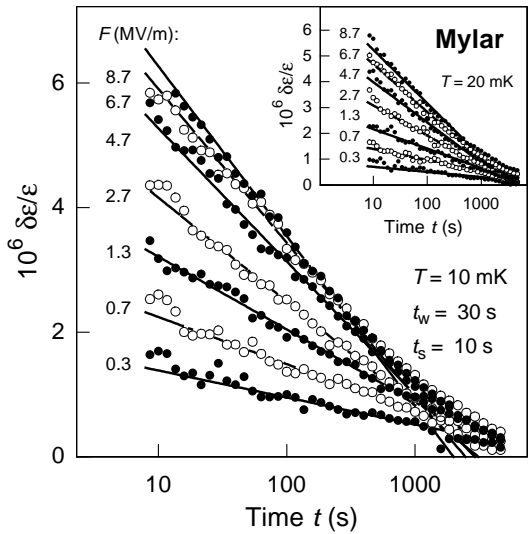}
\vskip -8mm 
\caption{Dielectric response of Mylar as in \fig{fig1}, but at constant sweep
times for several maximal bias fields $F$. The data are
taken at $T=10$~mK (main figure) and $T=20$~mK for the inset. Solid lines
are model curves as explained in the text.}
\label{fig2}
\end{figure}

Figs$.$~\ref{fig1} and~\ref{fig2} show the excess dielectric response of
Mylar after the temporary
application of an electric bias field for the waiting time $\tw=30$~s and at
temperatures of $T=10$~mK and $T=20$~mK.
The sketch in \fig{fig1} illustrates the evolution of the bias field sweep.
Note that the zero point of time is defined to be at the completion of the
sweep and, here, the time axes are logarithmic.
In \fig{fig1} the sweep times $\ts$ vary between
$0.3\,{\rm s}\le\ts\le300$~s at electric bias fields of maximal strength
$F=6.7$~MV/m, while in \fig{fig2} $\ts=10$~s for all curves at different
maximum bias fields $0.3\,{\rm MV/m}\le F\le8.7$~MV/m.

A temporary applied bias field causes relaxational processes that bring TSs out
of equilibrium. In~\cite{Lud03b}, strong deviations from the
expectations due to relaxation were observed in the dependence of the waiting
time (\fig{fig1} in~\cite{Lud03b}). These results were
qualitatively explained by considering the evolution of energy splittings
$E({\bf t})=\sqrt{\Delta_0^2+(\Delta+\pF(t))^2}$ while sweeping TSs through zero
asymmetry $\Delta+\pF(t)=0$ (assembling an avoided level crossing situation). If
the combination of electric ac-- and bias--field changes fast, systems tunneling
slowly cannot adjust to their momentary equilibrium while being symmetric,
\ie $\Delta+\pF(t)\lesssim\Delta_0$. Such TSs are driven non--adiabatically
out of equilibrium and, if part of a strongly coupled pair, suddenly
contribute to the dielectric response. Due to the range of relaxation times,
the excess response decays in time as $\log(\tau_0/t)$, as long as
the ac--field dominates the driving effect, \ie for small sweep rates $F/\ts$.
At faster sweep rates the bias--field drives TSs non--adiabatically. This
situation causes deviations from logarithmic decays~\cite{Lud03b,Lud03c}.
In \figs{fig1} and~\ref{fig2}, the transition between both non--adiabatic
limits can be seen~\cite{Lud03c}. Here, relaxational processes can be omitted,
as comparison to the waiting time dependence reported in~\cite{Lud03b}  
shows. 
The slopes of the solid lines in both figures, representing logarithmic decays
as expected for non--adiabatic ac--field driving,
are calculated from the results in~\cite{Lud03b}. They depend on temperature and
maximal bias field but not on sweep rate and waiting time~\cite{Lud03b}. 
\begin{figure}[th]
\includegraphics[width=0.95\linewidth]{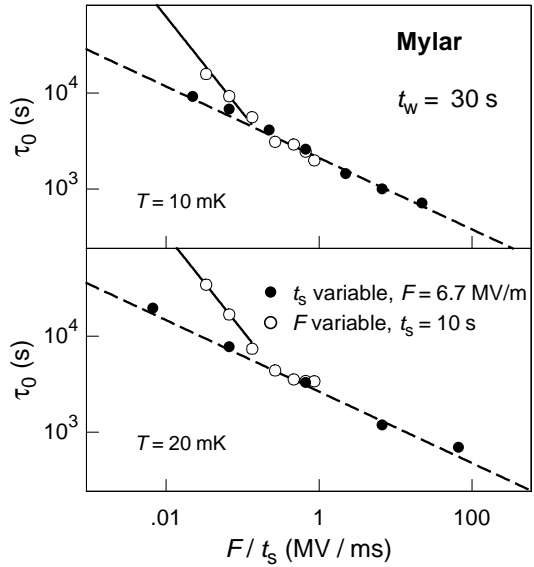}
\vskip -8mm 
\caption{Decay times for Mylar obtained from model curves of the previous
Figures due to a model including
dominant pair breaking by non--adiabatic driving of
TSs~\cite{Lud03b}. The solid lines are model curves without (and the dashed
lines with) an additional contribution to the dielectric response
that decays independent of temperature and bias field history.}
\label{fig3}
\end{figure}
The decay times $\tau_0$, defined as the times when the model curves extrapolate
to $\de=0$,
depend on the many parameters of the model, but we take them here as
fit parameters.

\Fig{fig3} shows double logarithmic plots of decay times as a function of the
sweep rate $F/\ts$. Solid points represent the decay times
extracted from measurements with a maximal bias field of $F=6.7$~MV/m and
different sweep times, for $T=10$~mK (\fig{fig1}) and $T=20$~mK in
the upper and lower plot, respectively. Non--adiabatic driving causes $\tau_0$
to be a function of the sweep rate $F/\ts$~\cite{Lud03b,Lud03c}. If
bias field driving is dominant (fast sweeps) the model introduced
in~\cite{Lud03b} proposes a
linear dependence $\tau_0=\xi(T)\ts/F$. For dominant ac--field
driving (slow sweeps) the exact form of the functional dependence is unknown.
All data sets presented in \fig{fig3} contain the transition between these two
limits (fast and slow sweeps). 
The decay times extracted from our sweep time dependent data (solid points)
show the same power law behavior over 4 orders of magnitude in $F/\ts$,
clearly indicating equal functional behavior in both limits.
However, we observe the relation $\tau_0\propto(\ts/F)^{0.4}$, in striking
contrast to the expected linear dependence.

The open circles in \fig{fig3} present the decay times obtained from the bias
field dependent measurements shown in \fig{fig2}.
For bias fields greater than a crossover field $F_{\rm c}\simeq1.5$~MV/m we
observe identical behavior as for the sweep time dependent measurements, that
were taken at $F=6.7$~MV/m, much greater than $F_{\rm c}$.
Strikingly, for $F<F_{\rm c}$ the behavior suddenly changes to the
expected linear relation $\tau_0\propto\ts/F$ (solid lines in \fig{fig3}).
The temperature dependence mirrors this situation:
For $F<F_{\rm c}$ we find $\tau_0(T=20 {\rm mK})/\tau_0(T=10 {\rm mK})\simeq1.8$
(solid lines), in agreement with the observed temperature
dependence of the decay times in the limit of dominant relaxational
processes~\cite{Lud03b}. For $F>F_{\rm c}$ we observe 
$\tau_0(T=20 {\rm mK})/\tau_0(T=10 {\rm mK})\simeq1.8^{0.4}$ (dashed lines),
with the identical exponent as for the sweep rate dependence in this limit.
Both the non--linear relation $\tau_0\propto(\ts/F)^{0.4}$ for $F>F_{\rm c}$ as
well as the temperature dependence cannot be explained by our model so far. 

\begin{figure}[th]
\includegraphics[width=0.95\linewidth]{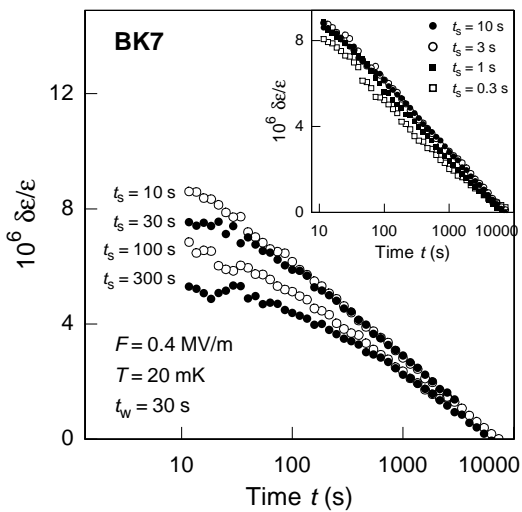}
\vskip -8mm
\caption{Sweep time dependence of the relative dielectric constant of BK7 after
transient application of a bias field $F=0.4$~MV/m at the temperature $T=20$~mK
for long sweep times $\ts\ge 10$~s in the main figure and short sweep times
$\ts\le 10$~s in the inset.}
\label{fig4}
\end{figure}

Our BK7--sample is twenty times thicker than the Mylar film resulting in an
accordingly weaker maximal bias field of $F=0.4$~MV/m for our maximum applied
voltage. However, the electric coupling between TSs is about 6 times stronger
than in Mylar~\cite{Rog97} yielding similar excess responses in both samples for
otherwise similar parameters. The sweep time dependence for BK7 is displayed in
the inset of \fig{fig4} for $\ts\le10$~s and in the main figure for
$\ts\ge10$~s.
Due to the stronger coupling between TSs in BK7, for
$\ts>\tw$ a sizable decay of the excess dielectric response already
happens while the bias field is still decreased back to zero.
In a simplifying approximation we assume an effective increase of the waiting
time proportional to the sweep time. For the data shown in \fig{fig4}, the linear
time transformation $t\rightarrow t+0.4\ts$ produces a set of almost identical
curves. This ability to compensate for the unwanted effect described above, is
gratifying but not essential for our analysis of the decay times, which are
almost unaffected by the above time transformation (due to the logarithmic
time scale).
The absence of any sweep time dependence in BK7 -- in striking contrast to the
observations in Mylar -- can be understood in terms of very different lower
cutoffs of the density of states for TSs, $\Emin$, in the two samples.
Only TSs with slow tunneling frequencies are driven non--adiabatically. Hence,
we expect to observe sweep rate independent decays, if $\Emin$ is larger than a
crossover splitting, $\Delta_{0\rm c}$, below which non--adiabatic driving of
TSs is present. Taking into account all of our
measurements~\cite{Lud03c}, we estimate $\Emin/\hbar\gtrsim 1.6$~mK for BK7 and
$\Emin/\hbar\lesssim 40~\mu$K for Mylar, in agreement with conclusions drawn
from a saturation of the dielectric constant at low temperatures~\cite{Rog97}.
For relaxational processes bringing pairs out of equilibrium, an upper limit for
the decay time can be estimated. By taking only one phonon
processes of TSs with $E\sim\kb T$ at zero field and $E\sim\pF\gg\kb T$ at
applied field into account and assuming a typical dipole moment of $p\sim1$~D
we find $\tau_0\lesssim\tw\pF/\kb T\sim100$~s. In contrast, we observe decay
times scattered around $\tau_0=7100\pm500$~s in our experiments (\fig{fig4}).
The excess dielectric response of BK7 is therefore primarily caused 
by a novel process, for which the decay time exhibits at most a very 
weak dependence on temperature~\cite{Lud03c}.

We will now show that the decay time dependence on the sweep rate
$\tau_0(F/\ts)$ in our Mylar sample
(\fig{fig3}) can as well be described by including such an additional
process with constant decay time. We assume that TSs contributing to the novel
process do not contribute to non--adiabatic driving and therefore
$\de\propto[1-\alpha(F)]\ln[\xi(T)\ts/(F t)]+
\alpha(F)\ln[\tilde\tau_0/t]$.
Here $\xi(T)\ts/F$ and $\tilde\tau_0$ are the decay times due to
non--adiabatic driving, containing the temperature dependent pre--factor
$\xi(T)$, and the novel process, respectively, and $\alpha(F)$ is the
amplitude of the novel process.
The decay times obtained from the logarithmic theory curves are then given by 
\be
\label{tau0}
\log\left(\tau_0\right)
      = \left[1-\alpha(F)\right]\log\left[\xi(T)\ts/F\right]
              +\alpha(F)\log\left[\tilde\tau_0\right]\;.
\ee
All lines in \fig{fig3} are obtained from~\eq{tau0}, where we
assumed $\alpha(F>F_{\rm c})=\alpha_0$ and $\alpha(F<F_{\rm c})=0$, \ie that
the novel process disappears for $F<F_{\rm c}$, in accordance with
experiments. The fit parameters are
$\tilde\tau_0\simeq4.7\cdot 10^3$~s, $\alpha_0\simeq0.6$ for both temperatures,
$\xi(10 {\rm mK})\simeq620$~MV/m and
$\xi(20 {\rm mK})/\xi(10 {\rm mK})\simeq 1.8$. By including the novel process we
can now recalculate the model curves for Figs$.$~\ref{fig1} and~\ref{fig2} as
well as for the data in a previous publication~\cite{Lud03b}. Hereby, the slopes
of all model curves only depend on two commen fit parameters, the dipole moment
of $p=1.2\pm0.1$~D and the interaction constant, $P_0U_0\simeq6.3\cdot10^{-4}$,
where $P_0$ is the density of states of TSs and $U_0$ the mean interaction
energy times volume. Note, that due to the influence of the novel process, our
value of $P_0U_0$, is slightly higher than previously reported~\cite{Lud03b}.
Using logarithmic decay curves for the novel process we find good agreement with
our data by assuming the same dependence of the slopes of the decay curves on
temperature and maximal bias field as already proposed in the "dipole gap"
theory~\cite{Bur95JLTP} and observed in our previous experiments for
relaxational pair breaking and non--adiabatic driving processes~\cite{Lud03b}.
Thus, the slope of the decay curves (in their logarithmic region) is
governed by the general dependence on temperature of the dielectric response as
well as how far the bias field drives the TSs out of equilibrium. 
In order to reach good agreement with all our measurements on
Mylar~\cite{Lud03c}, we necessarily must make the assumption in \eq{tau0} that
the novel process excludes additional contributions to the excess dielectric
response due to other processes, \ie non--adiabatic driving.
Undoubtedly, if the dynamics of strongly coupled pairs of TSs governs the other
contributions, the novel process influences clusters as well.

The temperature independence of $\tilde\tau_0$ points to a process involving
quantum mechanical tunneling, wherein thermal transitions are unimportant.
The general assumption of a flat distribution of tunneling parameters
$\rho(\lambda)=\rho_0$ in a disordered solid yields the distribution of
tunneling times $\rho(\tau_\lambda)=\rho_0/\tau_\lambda$, with
$\tau_\lambda\propto\exp(\lambda)$. Assuming that strongly coupled pairs stop
contributing to the dielectric response once they are in thermal equilibrium, as
in the "dipole gap" theory, we find a logarithmic decay
$\de\propto\int_{\tau_{\lambda,{\rm min}}}^{\tau_{\lambda,{\rm
max}}}\d\tau\rho(\tau)\Theta(t-\tau)=\rho_0\ln(\tilde\tau_0/t)$, with
$\tau_{\lambda,{\rm max}}\equiv\tilde\tau_0$ and the Heavyside step function
$\Theta(t-\tau)$.
A strong electric bias field couples to dipole moments, generally present in
amorphous solids due to displacements of atoms, and leads to
structural rearrangements of the atoms changing the potential energy
distribution. This situation yields a different set of low energy
states than the TSs provided at zero field. Nevertheless, in thermal
equilibrium, experimental observations are identical because of the glassy
distribution of parameters. Primarily, we are interested in the effect of
a bias field sweep on TSs contributing at zero field to the dielectric constant. 

Simulations suggest that the additional low energy internal degrees of freedom
observed in glasses include coherent movements of the order of 20--100
atoms~\cite{Schober}. The interaction of that many atoms yields a multiple
level scheme, wherein a TS can contribute to the excess dielectric response, 
if the two lowest levels are close enough in energy, \eg $\Delta_0\lesssim\kb T$
and $E\sim\kb T$. A strong bias field $\pF\gg\kb T$ couples to
dipole moments and reorganizes the energy eigenstates present in the glass. 
It thereby brings TSs, including strongly coupled pairs, out of thermal
equilibrium. Moreover, energy states that are thermally not accessible at zero
bias field will be occupied during the field sweep by quantum mechanical
tunneling. It is conceivable that successive tunneling processes while the bias
field is swept tend to bring TSs into metastable states that are lower in
energy than their neighbor potential wells (in configuration space) during the
entire field sweep, \ie tunneling processes tend to be unidirectional in
a non--equilibrium situation. Once back at zero bias field, these TSs are
generally still trapped in their metastable states and separated from
the lowest energy states (at zero bias field) by a large potential barrier.
Intermediate states temporarily occupied during the field sweep are now
statistically higher in energy and thermally not accessible
($\delta E\gg\kb T$).
While metastable states in single wells are occupied that cannot be counted as
TSs, additional metastable two level states exist. 
Because of the flat distribution of asymmetry energies, the overall
number of TSs contributing is the same at any time. However, TSs out of
equilibrium that belong to strongly coupled clusters contribute
to the excess dielectric constant as being isolated.
Our interpretation in terms of structural rearrangements should be considered as
a rough outline of a possible scenario. It calls for a detailed examination,
including the influence of metastable conditions on the dielectric response
dependent on the broadly distributed time constants of the different involved
processes.

We observe the novel process in BK7 at relatively small
bias fields. However, a minimum bias field is required
in the polyester glass Mylar possibly due to a potential barrier
necessary to rearrange hydrocarbon chains. 

In conclusion, after the temporary application of a sufficiently large electric
bias field to a glass sample at ultra low temperatures, we find a long--lasting
contribution to the excess dielectric response that
decays logarithmically in time with a decay time independent of temperature and
bias field history. We assign this novel process to field--induced structural
rearrangements in the glass that decay slowly by quantum mechanical tunneling
through large potential barriers. Strongly coupled pairs involved
cause an excess dielectric response as long as they are not in thermal
equilibrium.

We thank P$.$ Nalbach, J$.$ Baumgardner, J$.$ R$.$ Jameson, W$.$ Harrison
and S. Hunklinger for valuable discussions.
This work was supported by U$.$S$.$ Dept$.$ of Energy grant 
DE-FG03-90ER45435-M012 and the Deutsche Forschungsgemeinschaft.


\end{document}